\documentclass[conference]{IEEEtran}
\usepackage{amsmath,amsfonts}
\usepackage{amssymb}
\usepackage{enumerate,letltxmacro}
\usepackage{graphicx}
\usepackage{cite}
\usepackage{subfigure}
\usepackage{multirow}
\usepackage{algorithm}
\usepackage{algorithmic}
\usepackage{mathtools}
\usepackage{hyperref}
\usepackage{verbatim}
\usepackage{xcolor}

\hypersetup{draft}

\IEEEoverridecommandlockouts

\newtheorem{theorem}{Theorem}

\newtheorem{definition}{Definition}

\newtheorem{corollary}{Corollary}
\newtheorem{proposition}{Proposition}

\title{Undirected Unicast Network Capacity: \\ A Partition Bound}
\author{\IEEEauthorblockN{Satyajit Thakor and Mohammad Ishtiyaq Qureshi}
\IEEEauthorblockN{School of Computing and Electrical Engineering\\%, 
Indian Institute of Technology Mandi, Himachal Pradesh, India}
email: satyajit@iitmandi.ac.in, D15063@students.iitmandi.ac.in
}

\begin{document}
\maketitle
\begin{abstract}
In this paper, we present a new technique to obtain upper bounds on undirected unicast network information capacity. Using this technique, we characterize an upper bound, called partition bound, on the symmetric rate of information flow in undirected unicast networks and give an algorithm to compute it. Two classes of networks are presented for which the bound is tight and the capacity is achievable by routing thus confirming the undirected unicast conjecture for these classes of networks. We also show that the bound can be loose in general and present an approach to tighten it.
\end{abstract}
\section{Introduction}
Explicit characterization of the network information capacity, also called network coding capacity, is an open problem. Many outer bounds on the capacity \cite{AhlCai00,YanYan06,KraSav06,HarKle06,ThaGraCha09,KamTseAna11,ThaGraCha16a,KamAnaTseWan18} are known for directed acyclic multicast networks. However, there is limited progress on undirected unicast network information capacity problem. Even upper bounding the symmetric information rate on undirected unicast network information capacity is a challenging problem. Only two explicit upper bounds on symmetric information rate are known for general undirected networks: (1) the sparsity bound \cite{LeiRao99}, \cite{HarKle06} on symmetric rate is a trivial bound on both commodity and information flow and (2) the linear programming bound \cite[Chapter 15]{Yeu08}, \cite{HarKle06} using Shannon-type inequalities is generally not used for evaluation due to prohibitively large problem size.

Li and Li \cite{LiLi04} conjectured that, in undirected unicast networks, network coding cannot outperform routing in terms of the achievable rate. This conjecture is yet unsolved in general. In particular, it is known to hold for certain networks \cite{JaiVazYuv06,HarKle06,AlYon08} and certain classes of networks \cite{ForFul56,Hu63,OkaSey81,AlYon08,YinLiLiuWan17}. One approach to solve this conjecture is to obtain a characterization of the undirected unicast network information capacity and then check whether it matches the undirected unicast network routing capacity. However, this appears to be a difficult problem as only two simple upper bounds are known so far.

In this paper, we give a new upper bound, called partition bound, on the symmetric rate for information flow in general undirected unicast networks and an algorithm to compute it. We present partitioning technique to obtain upper bounds and prove tightness of the partition bound and the Li and Li's conjecture for two classes of networks. 

Section \ref{sec:prelim} provides some background on submodularity properties of entropy function, undirected unicast network model, some entropy inequalities and basic graph theory notions required in subsequent sections. In Section \ref{sec:Main Results}, we present the main results of the paper: a partition bound, an algorithm to compute the bound based on a recurrence relation, a partitioning technique, and tightness of the bound and proof of the Li and Li's conjecture for two classes of networks. In Section \ref{sec:Tightening the Partition Bound}, we show that the partition bound is not tight in general and also demonstrate an approach to tighten the bound. As a result, we present an alternative proof of the undirected unicast network capacity of Hu's 3-pairs network \cite{Hu63}.  The partition bound is tighter than the known bound \cite{AlYon08} for bipartite networks.

\section{Preliminaries}\label{sec:prelim}
\subsection{Submodularity of entropy}
\begin{definition}%[$(n,k)$-way submodularity]
For sets of random variables $A_1,\ldots,A_n$, define sets 
%\begin{align}
$$B_{i,n}\triangleq \bigcup_{\alpha\subseteq\{1,\ldots,n\}: |\alpha|=i} \left(\cap_{j\in \alpha} A_j\right)$$
%\end{align}
for $i=1,\ldots,n$. For $k\leq n$ the \textit{$(n,k)$-way submodulariy} is
\begin{align}
\sum_{i=1}^n h(A_i) \geq \sum_{i=1}^k h(B_{i,n})
\end{align}
where  $h(\cdot)$ is the Shannon entropy function.
\end{definition}

$n$-way submodularity for entropy function was proved in \cite{HarKle06}. $n$-way submodularity is equivalent to $(n,n)$-way submodularity. $(n,k)$-way submodularity for entropy function follows from $(n,n)$-way submodularity and non-negativity of entropy. Also note that $(n,k)$-way submodularity is equivalent to $(n,n)$-way submodularity if $B_{k+1,n}=\emptyset$ (which also implies $B_{i,n}=\emptyset$ for all $i>k+1$) since, by convention, we have $h(\emptyset)=0$.

\subsection{Network model}

In this work we focus on information capacity of undirected unicast networks. An undirected information network is denoted $G=(V,E,I,s,t)$ where $V$ is the set of nodes, $E$ is the set of edges of the form $e=\{u,v\}, u,v\in V$ and $I$ is the set of source indices with $|I|=k$. Mappings of a source to a node and a sink to a node are $s:I \mapsto V$ and $t:I \mapsto V$ respectively. In particular, source $i$ is located at node $s(i)$ and sink demanding $i$ is located at node $t(i)$. A network is unicast if a source located at a network node is demanded by exactly one sink located at a different network node. 

Also, for $V'\subseteq V$, $S(V')\triangleq \{i\in I: s(i)\in V'\}$ and $ST(V')\triangleq \{i\in I: s(i) \in V' \text{ or } t(i)\in V'\}$.  Now, consider the bi-directed 
version of the graph where each undirected edge $\{u,v\}$ is replaced with directed edges $(u,v)$ and $(v,u)$. For edge $e=(u,v)\in E$, denote  tail$(e)=u$ and head$(e)=v$. Consider disjoint subsets $V'$ and $V''$ of $V$.  The set of edges from nodes in $V'$ to nodes in $V''$ is denoted 
$${V'\rightarrow V''}\triangleq \{e\in E: \text{tail}(e)\in V', \text{head}(e)\in V''\}.$$ 
Similarly, ${V' \leftrightarrow V''}\triangleq ({V' \rightarrow V''}) \cup ({V'' \rightarrow V'})=(V' \cup V'') \rightarrow (V' \cup V'')$. For each source index $i$, we have associated source random variable $Y_i$. That is, $Y_i$ is available at $s(i)$ and is demanded at $t(i)$. The random variables flowing in a set of edges $V'\rightarrow V''$ is denoted $U_{V'\rightarrow V''}=(U_{e}:e\in V'\rightarrow V'')$. Also define the set of source indices $I(V',V'') \triangleq \{i: s(i)\in V', t(i) \in V''\}$.

By definition of network code \cite{Yeu08} (see also \cite{AhlCai00}), the edge variable $U_e$ is a function of the source random variables $Y_i, s(i)=\text{tail}(e)$ and edge random variables $U_{e'}, \text{head}(e')=\text{tail}(e)$. The decoding constraints are that each source random variable $Y_i$ is a function of $U_e, t(i)=\text{head}(e)$. It is assumed that the source random variables are mutually independent. Also, we have the unit capacity constraints $h(U_{u\rightarrow v})+h(U_{v\rightarrow u})\leq 1$ for each $\{u,v\}\in E$. Finally, an achievable rate tuple $(r_i:i\in I)$ must satisfy $r_i \leq h(Y_i)$ for all $i\in I$.

\begin{definition}%[symmetric \cite{AlYon08} or concurrent \cite{LeiRao99} rate]
For undirected network $G=(V,E,I,s,t)$, the \textit{symmetric \cite{AlYon08} or concurrent \cite{LeiRao99} rate}  of information flow is the scalar $r$ such that the tuple $(r_i=r: i\in I)$ is achievable.
\end{definition}

\subsection{Inequalities and $n$-partite graphs}
Following two well-known inequalities for the random variables involved in undirected network information are established in \cite{JaiVazYuv06}.

\begin{definition}%[$n$-partite graph]
An \textit{input-output inequality} \cite{JaiVazYuv06} for information flow for given $V'\subseteq V$ in $G=(V,E,I,s,t)$ is 
\begin{align}
h(Y_{ST(V')},U_{V'^c \leftrightarrow V'})\leq h(Y_{S(V')}, U_{V'^c \rightarrow V'})
\end{align}
where $V'^c \triangleq V \setminus V'$.
\end{definition}

Note that the input-output inequality is in fact a functional dependence relation induced by network coding constraints, i.e.,  $h(Y_{ST(v)},U_{V\setminus \{v\} \leftrightarrow v})= h(Y_{S(v)}, U_{V\setminus \{v\} \rightarrow v})$. However, viewing this functional dependence relation as the inequality, as the initial results in \cite{JaiVazYuv06} have suggested, can be useful to obtain an upper bound on information flow. 

\begin{definition}%[$n$-partite graph]
A \textit{crypto inequality} \cite{JaiVazYuv06} for information flow for given $V'\subseteq V$ in $G=(V,E,I,s,t)$ is 
\begin{align}
h(Y_{ST(V')\cap ST(V'^c)},U_{V' \leftrightarrow V'^c})\leq h(U_{V' \leftrightarrow V'^c}).
\end{align}
\end{definition}

Similar to sparsest cut bound \cite{LeiRao99} for multi-commodity flow, a sparsest cut bound \cite{HarKle06} for information flow in an undirected network follows from the crypto inequality. Now we describe a few basic notions for graphs.

\begin{definition}%[$n$-partite graph]
An undirected graph $G=(V,E)$ is \textit{$n$-partite} if $V$ can be partitioned into $n$ independent sets $P_1,\ldots, P_n$, where an independent set is a set of nodes such that there does not exist an edge between any pair of nodes in the set. $P=\{P_1,\ldots, P_n\}$ is called a\textit{ partition} and $P_i\in P$ is called a \textit{partition set} of $P$.
\end{definition}

Note that, if $G$ is $n$-partite then it is also $m$-partite for all natural numbers $m$ such that $n\leq m\leq |V|$.

\section{Main Results}\label{sec:Main Results}
Al-Bashabsheh \textit{et al}. \cite{AlYon08} gave an upper bound on symmetric information rate for bipartite undirected unicast networks. For clearer exposition of ideas, we first present a bound on symmetric information rate for 3-partite undirected networks, Proposition \ref{lem:1}, and then characterize a bound for general undirected unicast networks, Theorem \ref{thm:1}.

\subsection{A bound for 3-partite networks}
\begin{proposition}\label{lem:1}
For a 3-partite undirected network $G=(V,E,I,s,t)$, the symmetric rate of source information flow is upper bounded as
 \begin{align}
r \leq \frac{|E|}{|I|+|I(P_1,P_1)|+|I(P_2,P_2)|+|I(P_3,P_3)|}.\label{eq:1-1}
\end{align}
\end{proposition}

We present two proofs of the proposition. 
Proof 1 uses submodularity of entropy at an intermediate step (see \eqref{eq:5}) whereas Proof 2 uses $(n,2)$-way submodularity. While Proof 1 is lengthier, it will be instrumental in improving the bound in Section \ref{sec:Tightening the Partition Bound}. In contrast, Proof 2 is simpler but is not useful for improving the bound using a specific approach, however, it is easily extendable for general $n$-partite networks.
\begin{IEEEproof}[Proof 1 of Proposition 1]
Consider the bidirected version of the network. For a node $v\in P_1$, the input-output inequality is
\begin{align}
h(Y_{ST(v)},U_{P_2\cup P_3 \leftrightarrow v})&\leq h(Y_{S(v)}, U_{P_2 \cup P_3 \rightarrow v}).%\nonumber\\
%&\leq h(Y_{s(v)})+h( U_{P_2 \rightarrow v})+h(U_{P_3 \rightarrow v}) 
\label{eq:1}
\end{align}
Summing for all $v\in P_1$ and applying  $(n,2)$-way submodularity to obtain a lower bound on LHS, we get
\begin{align}
&h(Y_{ST(P_1)},U_{P_1 \leftrightarrow P_2\cup P_3})+h(Y_{I(P_1,P_1)})\nonumber\\
&\leq \sum_{v\in P_1} h(Y_{S(v)}, U_{P_2 \rightarrow v}, U_{P_3 \rightarrow v})\nonumber\\
&\leq \sum_{v\in P_1} h(Y_{S(v)})+ \sum_{e\in P_2\cup P_3 \rightarrow P_1}h(U_{e}).\label{eq:2}%\\
\end{align}
Similarly, for partitions $P_2$ and $P_3$, 
\begin{align}
&h(Y_{ST(P_2)},U_{P_2 \leftrightarrow P_1 \cup P_3})+h(Y_{I(P_2,P_2)})\nonumber\\
&\leq \sum_{v\in P_2} h(Y_{S(v)})+ \sum_{e\in P_1\cup P_3 \rightarrow P_2}h(U_{e})\label{eq:3}\\
&h(Y_{ST(P_3)},U_{P_3 \leftrightarrow P_1 \cup P_2})+h(Y_{I(P_3,P_3)})\nonumber\\
&\leq \sum_{v\in P_3} h(Y_{S(v)})+ \sum_{e\in P_1\cup P_2 \rightarrow P_3}h(U_{e}).\label{eq:4}
\end{align}
Now using the submodularity of entropy we have,
\begin{align}
&h(Y_I)+h(U_{P_1 \leftrightarrow P_2})\nonumber\\
&=h(Y_{ST(P_1)\cup ST(P_2)},U_{V \leftrightarrow V})+h(U_{P_1 \leftrightarrow P_2})\nonumber\\
&\leq h(Y_{ST(P_1)\cup ST(P_2)},U_{V \leftrightarrow V})+h(Y_{ST(P_1)\cap ST(P_2)}, U_{P_1 \leftrightarrow P_2})\nonumber\\
&\leq h(Y_{ST(P_1)},U_{P_1 \leftrightarrow P_2 \cup P_3})+h(Y_{ST(P_2)},U_{P_2 \leftrightarrow P_1 \cup P_3}).\label{eq:5}
\end{align}
Summing \eqref{eq:2} and \eqref{eq:3} and using the lower bound \eqref{eq:5},
\begin{align}
&h(Y_{I})+h(U_{P_1 \leftrightarrow P_2})+h(Y_{I(P_1,P_1)})+h(Y_{I(P_2,P_2)})\nonumber\\
&\leq \sum_{v\in P_1\cup P_2} h(Y_{S(v)})+ \sum_{e\in (P_2\cup P_3 \rightarrow P_1) \cup (P_1\cup P_3 \rightarrow P_2)}h(U_{e})\label{eq:6}
\end{align}
where we use the fact that $S(P_1),S(P_2)$ are disjoint since (a) source $i$ is available only at one network node and (b) $P_1$ and $P_2$ are disjoint. Also note that $(P_2\cup P_3 \rightarrow P_1)\cap (P_1\cup P_3 \rightarrow P_2)=\emptyset$.
Similarly, we can obtain
\begin{align}
&h(Y_{I})+h(U_{P_1 \leftrightarrow P_3})+h(Y_{I(P_1,P_1)})+h(Y_{I(P_3,P_3)})\nonumber\\
&\leq \sum_{v\in P_1\cup P_3} h(Y_{S(v)})+ \sum_{e\in (P_2\cup P_3 \rightarrow P_1) \cup (P_1\cup P_2 \rightarrow P_3)}h(U_{e})\label{eq:7}\\
&h(Y_{I})+h(U_{P_2 \leftrightarrow P_3})+h(Y_{I(P_2,P_2)})+h(Y_{I(P_3,P_3)})\nonumber\\
&\leq \sum_{v\in P_2\cup P_3} h(Y_{S(v)})+ \sum_{e\in (P_1\cup P_3 \rightarrow P_2) \cup (P_1\cup P_2 \rightarrow P_3)}h(U_{e}).\label{eq:8}
%
%&h(Y_{I})+h(U_{P_1 \rightarrow P_3, P_3 \rightarrow P_1})+h(Y_{I(P_1,P_1)})+h(Y_{I(P_3,P_3)})\nonumber\\
%&\leq \sum_{v\in P_1\cup P_3} h(Y_{s(v)})+ h(U_{P_2 \rightarrow P_1})+h(U_{P_3 \rightarrow P_1})+h(U_{P_1 \rightarrow P_3})+h(U_{P_2 \rightarrow P_3})\label{eq:7}\\
%&h(Y_{I})+h(U_{P_2 \rightarrow P_3, P_3 \rightarrow P_2})+h(Y_{I(P_2,P_2)})+h(Y_{I(P_3,P_3)})\nonumber\\
%&\leq \sum_{v\in P_2\cup P_3} h(Y_{s(v)})+ h(U_{P_1 \rightarrow P_2})+h(U_{P_3 \rightarrow P_2})+h(U_{P_1 \rightarrow P_3})+h(U_{P_2 \rightarrow P_3})\label{eq:8}
\end{align}
Now note that for terms $h(U_{P_1 \leftrightarrow P_2})$, $h(U_{P_1 \leftrightarrow P_3})$ and $h(U_{P_2 \leftrightarrow P_3})$ in LHS of \eqref{eq:6}, \eqref{eq:7} and \eqref{eq:8},
\begin{align}
h(Y_{I})&=h(U_{V \leftrightarrow V})\nonumber\\
&=h(U_{P_1 \leftrightarrow P_2},U_{P_1 \leftrightarrow P_3},U_{P_2 \leftrightarrow P_3})\nonumber\\
&\leq h(U_{P_1 \leftrightarrow P_2})+h(U_{P_1 \leftrightarrow P_3})+h(U_{P_2 \leftrightarrow P_3}).\label{eq:9}
\end{align}
Summing  \eqref{eq:6}, \eqref{eq:7} and \eqref{eq:8} 
\begin{align}
&3(Y_I)+\sum_{\{i,j\}\subset\{1,2,3\}} h(U_{P_i \leftrightarrow P_j})+2\sum_{i=1}^3h(Y_{I(P_i,P_i)}) \nonumber\\
&\leq 2 \sum_{v \in V} h(Y_{S(v)})+ 2\sum_{e\in V\leftrightarrow V}h(U_{e})\label{eq:9-1}
\end{align}
and applying  \eqref{eq:9}, we get
\begin{align}
%4h(Y_{I})+2\sum_{i=1}^3h(Y_{I(P_i,P_i)}) &\leq 2 \sum_{v \in V} h(Y_{S(v)})+ 2\sum_{e\in V\leftrightarrow V}h(U_{e})\label{eq:10}\\
4h(Y_{I})+2\sum_{i=1}^3h(Y_{I(P_i,P_i)}) &\leq 2 h(Y_I)+ 2|E|\\%\label{eq:10}\\
\Longrightarrow h(Y_{I})+\sum_{i=1}^3h(Y_{I(P_i,P_i)})&\leq |E|.\label{eq:11}
\end{align}
Hence a bound on the symmetric rate is
\begin{align}
r \leq \frac{|E|}{|I|+|I(P_1,P_1)|+|I(P_2,P_2)|+|I(P_3,P_3)|}\label{eq:11-1}
\end{align}
where, we used $r=r_i\leq h(Y_i)$ and source independence.
\end{IEEEproof}

\begin{IEEEproof}[Proof 2 of Proposition 1]
Summing \eqref{eq:2}-\eqref{eq:4} and applying $(n,2)$-way submodularity,
\begin{align*}
2h(Y_{I})+\sum_{i=1}^3h(Y_{I(P_i,P_i)})%\nonumber\\
&\leq \sum_{v\in V} h(Y_{S(v)})+ \sum_{e\in V\leftrightarrow V}h(U_{e})\nonumber\\
%\Longrightarrow 2h(Y_{I})+\sum_{i=1}^3h(Y_{I(P_i,P_i)})\nonumber\\
%&\leq h(Y_{I})+ |E|\\
\Longrightarrow h(Y_{I})+\sum_{i=1}^3h(Y_{I(P_i,P_i)})&\leq |E|\nonumber\\
\Longrightarrow r &\leq \frac{|E|}{|I|+\sum_{i=1}^3|I(P_i,P_i)|}.
\end{align*}
Note that, in this case $(n,2)$-way submodularity is equivalent to $(n,n)$-way submodularity.
\end{IEEEproof}

\subsection{A bound for general networks}

\begin{theorem}[Partition bound]\label{thm:1}
For an undirected network $G=(V,E,I,s,t)$, the symmetric rate of information flow is upper bounded as
 \begin{align}
r &\leq \min_{P} \frac{|E|}{|I|+\sum_{i=1}^{n}|I(P_i,P_i)|}\nonumber\\
&= \frac{|E|}{|I|+\max_{P} \sum_{i=1}^{n}|I(P_i,P_i)|}%\nonumber%\label{eq:1-1}
\end{align}
where $P$ is a partition of $V$ into independent sets $P_1,\ldots,P_{n}$.
\end{theorem}

\begin{IEEEproof} 
The proof is similar to Proof 2 of Proposition 1 for the 3-partite case. Consider a valid partition $P=\{P_1,\ldots,P_n\}$. For a partition set $P_i\in P$, we have
 \begin{align}
 &h(Y_{ST(P_i)},U_{P_i \leftrightarrow \cup_{j\neq i}P_j})+h(Y_{I(P_i,P_i)}) \nonumber\\
 &\leq \sum_{v\in P_i} h(Y_{S(v)})+ \sum_{e\in \cup_{j\neq i}P_j \rightarrow P_i}h(U_{e}).%\label{eq:2}%\\
 \end{align}
 Now, summing such inequalities for all $i$'s and applying $(n,2)$-way submodularity (which is equivalent to $(n,n)$-way submodularity for these sets), 
 \begin{align}
&h(Y_{I})+h(U_{V\leftrightarrow V})+\sum_{i=1}^nh(Y_{I(P_i,P_i)})\nonumber\\
&\leq \sum_{v\in V} h(Y_{S(v)})+ |E|\\
\Longrightarrow&2h(Y_{I})+\sum_{i=1}^nh(Y_{I(P_i,P_i)})%\nonumber\\
\leq h(Y_{I})+ |E|\\
%\Longrightarrow2h(Y_{I})+\sum_{i=1}^{|I^*|}h(Y_{I(P_i,P_i)})%\nonumber\\
%&\leq \sum_{v\in V} h(Y_{s(v)})+ |E|\\
\Longrightarrow& r \leq  \frac{|E|}{|I|+\sum_{i=1}^{n}|I(P_i,P_i)|}.%\label{eq:1-1} 
 \end{align}
\end{IEEEproof}

The results so far suggest a technique for obtaining upper bounds on undirected unicast network capacity. We state this technique, called \textit{partitioning technique}, explicitly as follows:
\begin{description}
\item[\textit{Step 1}: ] \ Consider a partition of nodes into independent sets.
\item[\textit{Step 2}: ] \ For each independent set obtain an information inequality using input-output inequalities for the nodes in the independent set and submodularity.
\item[\textit{Step 3}: ] \ Combine thus obtained inequalities for independent sets to obtain an upper bound.
\end{description}

Using this technique we have obtained the partition bound and we will use the same basic technique together with functional dependence relations to obtain a tighter bound for a 3-pairs network  in Section \ref{sec:Tightening the Partition Bound}.
\subsection{Computing the partition bound}
Theorem 1 suggests that, to evaluate the bound, it is sufficient to consider a partition such that the total number of source-sink pairs in a same partition set is maximized. 
Here we describe a way of finding an optimal partition via establishing a recurrence relation. 

Let $P^*$ be an optimal partition and opt$(I)$ be a biggest subset of $I$ such that for all $i \in  \text{opt}(I)$ we have $\{s(k),t(k)\}\subseteq P_i$ for some $P_i\in P^*$. Note that for some $i\in I$ if $\{s(i),t(i)\}\in E$ (considering undirected version) then $i$ cannot be in opt$(I)$ and hence we can restrict the search for opt$(I)$ in the set
\begin{align}
\hat I \triangleq \{i\in I: \{s(i),t(i)\}\not\in E\}.
\end{align}
Now, let the set of neighboring nodes of  $u$ be $\text{ne}(u)=\{v\in V: \{v,u\}\in E\}$. There are two possibilities for any $k\in \hat I$:
\begin{enumerate}
\item $\{s(k),t(k)\}\subseteq P_i$ for some $P_i$ then $\text{opt}(\hat I)=\{k\}\cup \text{opt}(\hat I \setminus(\text{conf}(k)\cup\{k\}))$, where conf$(k)$ (abbreviated conflicting subset) is
\begin{align*}
\text{conf}(k)\triangleq &\{l\in \hat I : [t(l)\in \{s(k),t(k)\},s(l)\in\text{ne}(s(k))] \\
&\text{ or }[s(l)\in\{s(k),t(k)\}, t(l)\in\text{ne}(s(k))]\}.
\end{align*}
\item $\{s(k),t(k)\}\not\subseteq P_i$ for any $P_i$ then $\text{opt}(\hat I)=\text{opt}(\hat I\setminus \{k\})$.
\end{enumerate}  
Hence we have a recursive formula
\begin{align*}
&|\text{opt}(I)|=|\text{opt}(\hat I)|\nonumber\\
&=\max \left\{|\text{opt}(\hat I\setminus \{k\})|, 1+ |\text{opt}(\hat I\setminus(\text{conf}(k)\cup\{k\}))|\right\}.
\end{align*}

The value $|\text{opt}(I)|$ for a given network can be computed using Algorithm \ref{algo:Recursive}. Following this discussion, we can recast the partition bound as follows.
\begin{corollary}
For an undirected network $G=(V,E,I,s,t)$, the symmetric rate of information flow is upper bounded as
 \begin{align}
r \leq \frac{|E|}{|I|+|\text{opt}(\hat I)|}.%\nonumber%\label{eq:1-1}
\end{align}
\end{corollary}

\begin{algorithm}
\caption{{$\textbf{Opt}(G,\hat{I})$}} \label{algo:Recursive}
  \begin{algorithmic}[1]
    \REQUIRE ${G}, \hat{I}$
    \ENSURE{$|\text{opt}(\hat{I})|$}
    \IF{$|\hat{I}|\in\{0,1\}$}
    \RETURN {$|\hat{I}|$}
    \ELSE
    \RETURN{$\max \{\textbf{Opt}(G,\hat I\setminus \{k\}),    1+\textbf{Opt}(G, \hat I\setminus(\text{conf}(k)\cup\{k\}))\}\quad \quad\backslash\backslash$ Here, choose any $k\in\hat{I}$. }
    \ENDIF
  \end{algorithmic}
\end{algorithm}

\subsection{The capacity for some classes of networks}

Note that the partition bound is equivalent to the bound \cite[Theorem 3]{AlYon08} and is tight for the complete bipartite network $K_{3,2}$ \cite{JaiVazYuv06}. 
In \cite{AlYon08}, Type-I and Type-II bipartite networks are defined for which the Li and Li conjecture was established.
As a generalization, now we present two classes of $n$-partite networks for which the partition bound is tight and attainable by a routing scheme. Thus, the Li and Li's conjecture is established for these classes networks.

\begin{definition}%[Type-I and Type-2 $n$-partite networks]
\textit{Type-I $n$-partite network} is a complete $n$-partite network such that for every unordered pairs of nodes in a partition, there is a source-sink pair and there are no other source-sink pairs. \textit{Type-II $n$-partite network} is a complete $n$-partite network such that for every unordered pairs of nodes in the network there is a source-sink pair and there are no other source-sink pairs.
\end{definition}

\begin{proposition}
For Type-I and Type-II $n$-partite networks, the partition bound is tight and attainable by a routing scheme.
\end{proposition}
\begin{IEEEproof}[Proof (sketch)]
The partition bound on the symmetric rate for Type-I networks is ${|E|}/{2\sum_{i=1}^n I(P_i,P_i)}.$ 
Now note that, if each edge has capacity $$2\sum_{i=1}^n I(P_i,P_i)=\sum_{i=1}^{n}|P_i|(|P_i|-1)$$ then we can attain the symmetric rate of $\prod_{i=1}^{n}|P_i|$ by the routing scheme which delivers information from a source to a sink in two hops. 
But, by assumption, each edge has unit capacity and hence the routing scheme can attain the symmetric rate $|E|/(2(\sum_{i=1}^n I(P_i,P_i)))$ where $|E|=\prod_{i=1}^{n}|P_i|$.

The partition bound on the symmetric rate for Type-II networks is $$\frac{|E|}{\sum_{\{i,j\}\subseteq \{1,\ldots,n\}: i\neq j}I(P_i,P_j)+2\sum_{i=1}^n I(P_i,P_i)}.$$ Now note that, if each edge has capacity 
\begin{align*}
&\sum_{\{i,j\}\subseteq \{1,\ldots,n\}: i\neq j}I(P_i,P_j)+2\sum_{i=1}^n I(P_i,P_i)\\
&=\prod_{i=1}^{n}|P_i|+\sum_{i=1}^{n}|P_i|(|P_i|-1)
\end{align*} 
then we can attain the symmetric rate of $\prod_{i=1}^{n}|P_i|$ by the routing scheme which is obtained by considering the routing scheme for Type-I networks and then superpositioning the flow of $\prod_{i=1}^{n}|P_i|$ for source-sink pairs which are one hop away. But each edge has unit capacity and hence the routing scheme can attain the symmetric rate $|E|/(\prod_{i=1}^{n}|P_i|+2(\sum_{i=1}^n I(P_i,P_i)))$ where $|E|=\prod_{i=1}^{n}|P_i|$.
\end{IEEEproof}

\section{Tightening the Partition Bound}\label{sec:Tightening the Partition Bound}
\begin{figure}[htbp]
\centering
  \includegraphics[scale=.65]{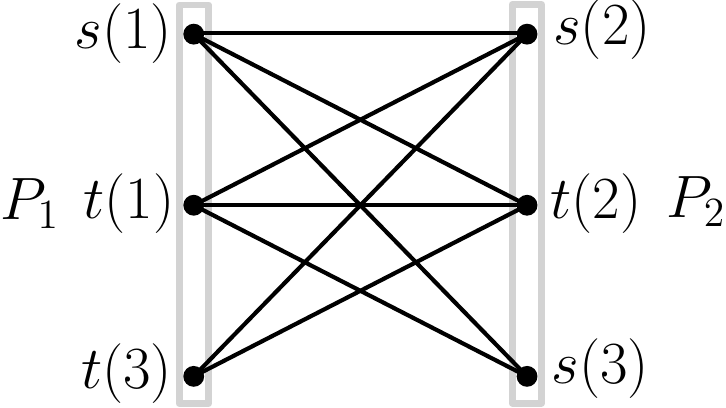}
  \caption{Hu's 3-pairs bipartite network.}\label{fig:HuNet1}
\end{figure}

%\subsection{Hu's 3-pairs Network}
Hu's 3-pairs network is bipartite and is depicted in Figure \ref{fig:HuNet1}. The bound by Al-Bashabsheh and Yongacoglu  \cite[Theorem 3]{AlYon08} for this bipartite network evaluates to $8/5$. 

\begin{figure}[htbp]
\centering
  \includegraphics[scale=.65]{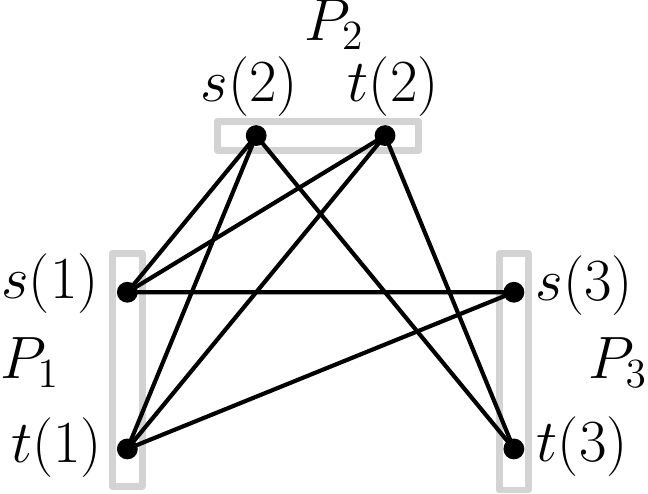}
  \caption{Hu's 3-pairs network with three partitions.}\label{fig:HuNet}
\end{figure}

Figure \ref{fig:HuNet} shows the Hu's 3-pairs network \cite{Hu63} with a particular partition of nodes into three independent sets. The partition bound is $8/6$ which is the same as the sparsity bound  \cite{LeiRao99}, \cite{HarKle06} but the information flow capacity (and commodity flow capacity too)  is $8/7$ \cite[Theorem 2]{AlYon08}. Thus, the partition bound is loose in general. However, the Hu's network demonstrates that, for a bipartite network, the partition bound can be strictly tighter than the bound given in \cite[Theorem 3]{AlYon08}. Also note that \cite[Theorem 3]{AlYon08} is implied by (or is a special case of) the partition bound.

In the following, we tighten the partition bound for Hu's network to its information capacity 
and thus present an alternative and simpler proof of \cite[Theorem 2]{AlYon08}.
 \begin{proposition}
 The symmetric rate of information flow in Hu's network is  at most $8/7$.
 \end{proposition}
\begin{IEEEproof}
We tighten the partition bound by making a modification in Proof 1 of Proposition \ref{lem:1} for Hu's network. Consider the partition $\{P_1,P_2,P_3\}$ as shown in Figure \ref{fig:HuNet}. Now consider the terms $h(U_{P_1 \leftrightarrow P_2})$, $h(U_{P_1 \leftrightarrow P_3})$ and $h(U_{P_2 \leftrightarrow P_3})$ in LHS of \eqref{eq:6}, \eqref{eq:7} and \eqref{eq:8}. Note that, $P_i \leftrightarrow P_j$ for each $\{i,j\}\subset\{1,2,3\}$ separates $s_3$ and $t_3$ and thus (as a consequence of the crypto inequality)
\begin{align}
h(Y_3|U_{P_i \leftrightarrow P_j})=0, \{i,j\}\subset\{1,2,3\}.
\end{align}
This implies
\begin{align}
&\sum_{\{i,j\}\subset\{1,2,3\}} h(U_{P_i \leftrightarrow P_j})\nonumber\\
&=h(Y_3, U_{P_1 \leftrightarrow P_2})+h(Y_3, U_{P_1 \leftrightarrow P_3})+h(Y_3, U_{P_2 \leftrightarrow P_3})\nonumber\\
&\geq h(Y_3)+h(Y_3)+h(U_{V \leftrightarrow V})
\end{align}
by submodularity of entropy. Using this lower bound on $\sum_{\{i,j\}\subset\{1,2,3\}} h(U_{P_i \leftrightarrow P_j})$ in \eqref{eq:9-1}, that is,
\begin{align}
&3(Y_I)+\sum_{\{i,j\}\subset\{1,2,3\}} h(U_{P_i \leftrightarrow P_j})+2\sum_{i=1}^3h(Y_{I(P_i,P_i)}) \nonumber\\
& \leq 2 \sum_{v \in V} h(Y_{S(v)})+ 2\sum_{e\in V\leftrightarrow V}h(U_{e})\\
\Longrightarrow &4h(Y_{I})+2h(Y_3)+2\sum_{i=1}^3h(Y_{I(P_i,P_i)}) \nonumber\\
&\leq 2 \sum_{v \in V} h(Y_{S(v)})+ 2\sum_{e\in V\leftrightarrow V}h(U_{e})\\
\Longrightarrow &h(Y_{I})+h(Y_3)+\sum_{i=1}^3h(Y_{I(P_i,P_i)})\leq |E|\\
\Longrightarrow &7r \leq 8%\frac{8}{7}
\end{align}
which can be attained by a routing scheme.
\end{IEEEproof}

Note that, though Proof 2 of Proposition \ref{lem:1} is simpler, a similar modification cannot be made in it for obtaining a tighter bound for Hu's network.

\section{Conclusion}
We characterized a partition bound on the symmetric rate of undirected unicast network capacity. Two proof methods were presented. The bound was obtained by partitioning the set of vertices into independent sets and then applying information inequality constraints in a certain way to obtain a converse type result for the capacity problem. A recurrence formula was established for computing the partition bound. Two classes of networks were described for which the partition bound is tight and this bound on the symmetric rate can be attained by routing thus proving the Li and Li's conjecture for these classes of networks. Finally, we showed that the partition bound is not tight in general and demonstrated a tight bound for Hu's 3-pairs network via modification in the proof of the partition bound.
\section*{Acknowledgment}
This work is supported by SERB, Department of Science and Technology, Government of India, under Extra Mural Scheme SB/S3/EECE/265/2016. 
\bibliographystyle{ieeetr}
\bibliography{network}
\end{document}